# Design and Testing for Steel Support Axial Force Servo System

Sana Ullah[1], Yonghong Zhou [2], Maokai Lai[1], Xiang Dong [1], Tao Li[1], Xiaoxue Xu[1], Yuan LI[1], Ting PENG[1*]

1- Key Laboratory of Highway Engineering Ministry of Education in Special Areas, Chang'an University, Xi'an, 710064

2- Sichuan Chuanjiao Road and Bridge Co., LTD, Guanghan, Deyang, Sichuan 618300

Corresponding Author: t.peng@ieee.org

**Abstract—** Foundation excavations are deepening, expanding, and approaching structures. Steel supports measure and manage axial force. The study regulates steel support structure power during deep excavation using a novel axial force management system for safety, efficiency, and structural integrity. Closed-loop control changes actuator output to maintain axial force based on force. In deep excavation, the servo system regulates unstable soil, side pressure, and structural demands. Modern engineering and tech are used. Temperature changes automatically adjust the jack to maintain axial force. Includes hydraulic jacks, triple-acting cylinders, temperature, and deformation sensors, and automatic control. Foundation pit excavation is dynamic, yet structure tension is constant. There is no scientific way to regulate axial force foundation pit excavation. The revolutionary Servo system adjusts temperature, compression, and axial force to deform pits. System control requires foundation pit direction detection and modification. This engineering method has performed effectively for deep foundation pit excavation at railway crossings and other infrastructure projects. The surrounding protective structure may reduce the steel support's axial stress, making deep foundation excavation safe and efficient.

*Keywords— Servo systems, Steel strut support design, Deformation control, Monitoring and control methods, Deep excavation projects.*

## I. INTRODUCTION:

Chinese urban train development puts station foundations at risk: Like Hanzhong Road station on Shanghai Metro Line No. 13, transfer station pit excavation was 33 meters. Two city center foundation holes' approach buildings [28, 35]. The Wuding Road station pit of Shanghai Metro Line No. 14 is 2 meters from neighboring buildings. Environmental deformation limits [19, 20]. Before Shanghai Airport Express massive foundation excavation, HSR. HSR may move 2 mm, minimum 10.6 meters, in foundation holes. Deformation in soft soil foundation pits is considerable and difficult to control. Solutions must work [3, 9, 21]. Increasing retaining structure stiffness, strut spacing, and pre-stress minimizes foundation pit deformation [1, 13, 14, 18]. In soft soil excavations, servo steel strut support systems are employed increasingly to control deformation. Figure 1. Shanghai and Ningbo Metros feature 30 and 20 mechanical steel-strut-supported station foundation holes. An integrated hydraulic jack allows servo steel struts actively control wall deformation by adjusting axial force or length [4, 17]. Adjustable axial force foundation pit wall deformation. Strut is hydraulically elevated with little axial force. Wall deflection may surpass estimates, requiring deformation mitigation. Adjust struts. Hydraulic pressure builds till built-in jack is long enough. Servo steel struts don't level bases. Ningbo project foundation pit's servo steel strut system failed to stop 16-m 110 mm lateral wall displacement. Servo steel strut design and use lack axial force magnitude criterion. Top site engineers detected excavation deformation control difficulties. Controlling servo strut deformation was examined. Di et al. [7] compared servo and regular steel struts for deformation control using real-world data. Servo steel struts may reduce deformation and axial force losses, the research revealed. Huang and colleagues [15] used a spring-beam model to reproduce servo steel struts and assess wall deformation. Mixture boundary circumstances have spring and focused forces. Li et al. [17] controlled servo steel strut deformation using axial tensions. Servo steel strut performance was simplified by removing excavation-related active axial force adjustment. Axial force adjustment must be realistic to maximize servo steel strut use. Measure lateral earth pressure from servo strut axial forces by deforming diaphragm wall. Wall deformation and lateral earth pressure mobilization aid structural design. Not linearly affected by wall deformation. Over manipulating strut axial force impacts rearward diaphragm walls and ground pressure. Previous research [11, 12] confirm this. Lateral earth pressure is limited by strut deep stress [5, 10]. First- and second-generation steel support axial force servo systems changed foundation pit engineering by improving excavation stability, control, and safety. These innovative techniques restrict designs yet control excavation movement and deep excavation project structural integrity. The new control servo system design and test measurements of the steel support axial force servo system adjustment device used to construct a deep foundation hole in the Sichuan Chuangye Avenue West Extension (East Section) channel are detailed.

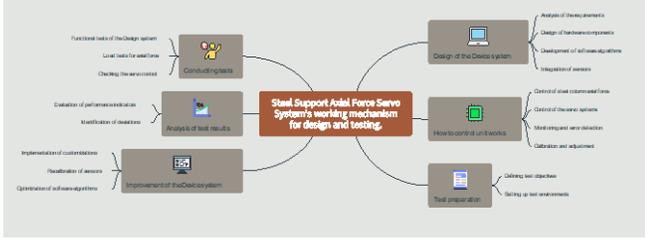

*Fig.1: Working mechanism Flowchart of servo system*

Develop and test a steel support axial force servo control system. A precise and stable control system to regulate axial force, optimize its fast response time and high accuracy, ensure the steel support structure's safety and reliability using an efficient control mechanism, test its performance under various operating conditions, and provide practical insights for its implementation and maintenance are the project's goals.

## II. AXIAL FORCE COMPENSATION TECHNOLOGY OF STEEL SUPPORT:

The axial force servo system excavated large foundation holes around railway crossings and other vital infrastructure. Steel stabilizes axial force. Most system control is hydraulic and automated. The utility model actively manages foundation pit retaining structure deformation using surrounding protective object deformation needs. Active compression and controlled deformation result from deep foundation pit steel support axial force. Expected strong foundation excavation protection.

Construction points:

1. 24/7 foundation pit steel servo axial force monitoring.
2. Tripled-acting cylinder, three-way automatic adjustment.
3. The bearing sleeve box steel force transfer support secures the hydraulic cylinder for real-time axial force adjustment.

## III. SUPPORT AXIAL FORCE CONTROL SERVO SYSTEM:

Advanced control mechanisms regulate foundation pit horizontal displacement in the Supporting Axial Force Control Servo System. Main components include hardware and software. Engineering applications needing accurate excavation-induced foundation pit side wall deformation. The method regulates foundation pit side wall deformation 24/7. High-pressure alarms and low-pressure servos prevent foundation pits. The foundation pit excavation protects the iron tower. Sichuan's West Extension Line of Chuangye Avenue (East Section) steel support structure uses controls system. CNC pump station, support head, and main engine are in Figure 2.

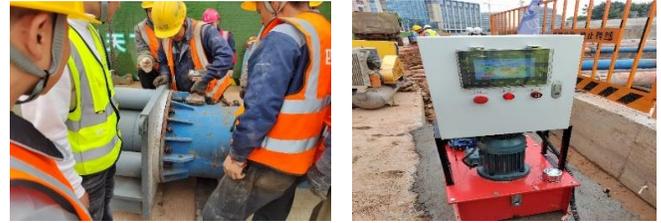

*Fig.2: Main Parts of the Control Servo systems*

## IV. STRUCTURE OF SERVO SYSTEM FOR STEEL SUPPORTING AXIAL FORCE:

Many applications use servo systems for precise positioning and force control. Axial support. A strong servo framework is needed. Mechanical locks support head. Enhanced Redundancy Three locks improve fault tolerance. The other two locks may prevent collapse if one fails. Important because system or structural failures may be severe. Improved load distribution. Three equally spaced locks adjust head tension. This reduces lock stress, improving system stability and component life. More Tolerance for Mechanical parts break. Without repairing three locks, two may break. This reduces system downtime. Its shows tensile testing of material.

### A. System Development:

High-strength steel support beams, cross-bracing components, adjustable hydraulic jacks, and sensor technologies improve deep excavation stability and safety.

- Project department control room main control system and display.
- According to foundation pit design, the numerical control hydraulic pump station is positioned.
- Automatic compensation end for steel-supported foundation pit side wall.
- Mainframe software-based management platform on display device.
- Lateral extraction pushes in. Auto-compensating steel supports adjust horizontal displacement for users. Figure 3 shows the automated axial force compensating steel support system's components.

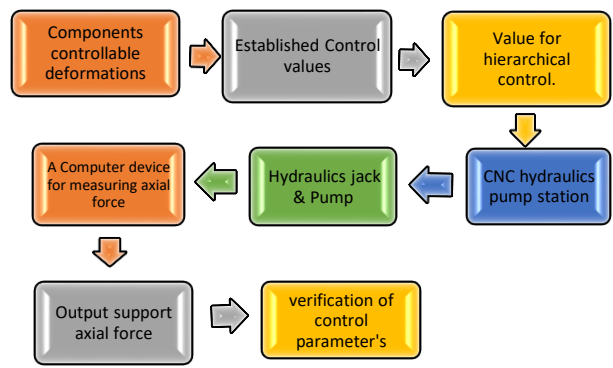

*Fig.3: Servo Working Flowchart System steel strut designed to support axial force*

### B. *Working principle:*

The system controls essential components. Weight-distributing steel cross-girders. Sensors monitor deformation and temperature to automate performance improvements, while hydraulic jacks manage axial stresses. The working servo method permits axial force in Fig 4.

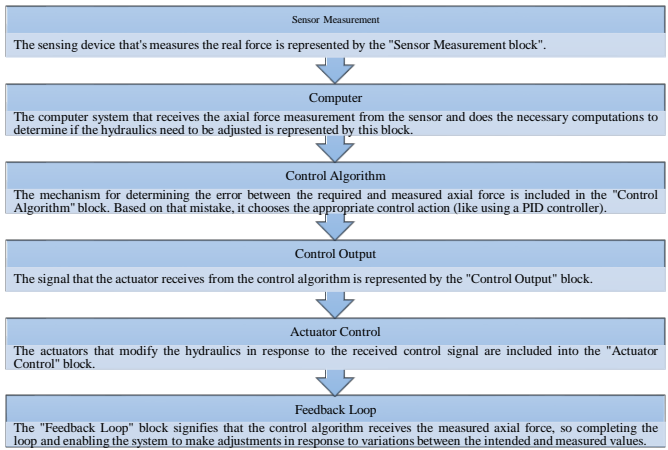

*Fig.4: Block diagram visually represents the Control loop (showing the interconnected components & flow of the information)*

How the control loop works PC detects axial force?

Computers forecast and decrease axial force failures. The actuator regulates pump and jack hydraulics. The sensor sends computer axial force data for adjustment. It persists.

Steel-supported axial-force close-loop servos. Controller assists steel. Strain gauges and load cells measure structure. A PID controller. Controller error messages are calculated using sensor force feedback and anticipated force set point. Steel support structure force is insufficient in erroneous signals. Data difficulties cause controllers to alert actuators. Pneumatic actuators, hydraulic cylinders, and screw motors control steel support structure force. Controlling actuator output retains structural force and eliminates false signals. The servo system compares force to set point, creates an error signal, and alters actuator output to deliver steel support structure axial force. Live closed-loop system control stabilizes the structure.

### C. *Configuration of the System:*

The essential hardware elements of the servo system that controls axial force on a steel support structure and their force management tasks are briefly described in Table 1. Hardware for servos includes:

- **hydraulic jack:** The hydraulic jack is the main source of axial force. It has pistons and cylinders.
- **hydraulic power unit (HPU):** The hydraulic power unit (HPU) supplies pressurized hydraulic fluid to the hydraulic jack. include valves, reservoirs, pumps, and motors.
- **Control System:** Regulates hydraulic fluid flow, pressure, and direction to ensure system functionality.
- **Sensors:** Use sensors to detect factors including axial force, hydraulic pressure, and piston position for feedback control. Measure foundation pit temperature and deformation.
- **Actuators:** To manage pit deformation, adjust the supporting axial force using actuators.
- **Control Unit:** The Control Unit processes sensor data and creates actuator orders.

*Table 1 Configuration of the Servo system*

| S/No. | Component | Description |
|---|---|---|
| 1 | Actuator | Hydraulic/Pneumatic Actuator or Electric Motor |
| 2 | Force Sensor | Load Cells or Strain Gauges |
| 3 | Controller | PID (Proportional-Integral-Derivative )Controller for force regulation |
| 4 | Power Supply | Provides electrical power to system components |
| 5 | Amplifier | Boosts control signal output for the actuator |
| 6 | Feedback mechanism | Monitors actual force and adjusts the control signal |
| 7 | Safety features | Limit switches, Emergency stop buttons, Overcurrent protection |

Control force Servo head assemblies with three locks are safer, more precise, more dependable. highlights this setup's benefits:

- Triple-lock systems are redundant, low-capacity, and pressure-consistent. Preventing bridge and heavy

equipment system failure during long downtime or collapse. Reduced capacity allows preventive maintenance and repairs, preventing partial lock failures. Stable three-lock assemblies distribute load uniformly. Load distribution and framework-servo connection help head assembly.

- Wear tolerance, operation, and downtime increase with three-lock systems. For preventative maintenance, sensors and locking mechanisms monitor lock engagement and wear. Cheap, stronger-than-double locks save head assembly space.
- Application and risk tolerance decide two- or three-lock designs. Three-lock systems are accurate and dependable for high-risk, reliable applications, whereas double-lock systems may be cheaper for low-risk, low-downtime applications.

### D. Hardware composition

The shaft force automated compensation steel support system hardware contains the main control system, numerical control hydraulic pump station, and automatic compensation end. From the numerical control hydraulic pump station, the main control system and automated compensation end measure and regulate steel supporting axis force. Closed-loop. Controlling steel structural support force. Structure strain and load gauges' measure force. PID controls systems. Controllers detect faults by comparing sensor feedback force to force set point. The error signal indicates inappropriate steel support structural force. Accidental actuator control. The actuator regulates steel support framework force. Motor, actuator, or hydraulic cylinder screwed. Actuator output adjustment decreases signal mistakes and maintains structural force. After comparing force to the set point, the servo system delivers an error signal, modifies actuator output, and applies axial force to the steel support structure.

*Fig.5: CNC pump station power connection & oil filling*

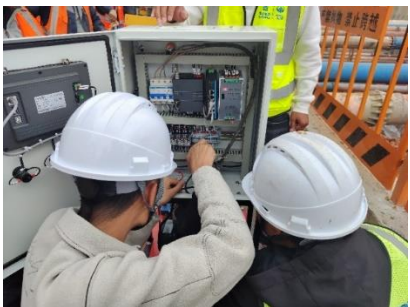

### E. Installation process

Figure 6 shows steel purlin installation, ground support head splicing, lifting, pre-stressing, and real-time axial force monitoring.

1. Station near HV. Used 10m steel pipe purlin. All-welded steel four-beam Triangulum carrier. Cement mortar supports steel purlin and diaphragm.
2. Ground steel pipe head/purlin. Correct steel head bolt spacing. Big head prep after Assembly. Bolts and steel plate secure the support head.
3. Cranes raise the steel support near the high-pressure tower, gently drop it on the steel purlin, and manually adjust it to design.
4. Instantly connect steel support, data pump station, head oil pipeline, and pressure.

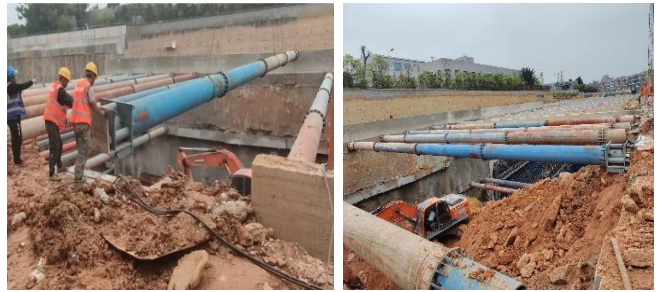

*Fig.6: installation process*

## V. REAL-TIME MONITORING

Real-time foundation hole drilling requires automation. Engineering construction impacts foundation pit structure and protection, enabling dynamic control and information creation. Early warning, process management, and construction safety improve with data monitoring (Fig. 7).

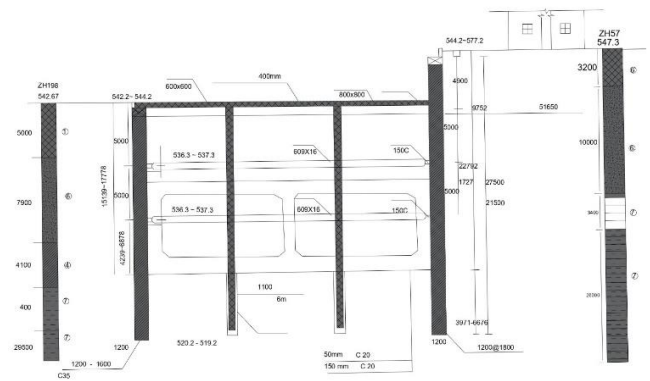

*Fig.7: Sectional schematic diagram of the supporting structure and monitoring location*

Ultrasonic and pressure sensors measure steel support head assembly axial force and displacement. Specific axial force and dynamic range activate the mechanism immediately. Adjusts automatically if steel support axial force falls below design. The steel support lowers automatically if its axial force exceeds the limit. Less foundation excavation distortion and high-voltage line sinking. The monitoring end's axial force may be manually adjusted to match shaft force in various operations. Value may change if data reaches early detection threshold. Title color and unprocessed alert alarm levels appear after automatic Alarm Centre cleaning.

*A. Monitoring and control methods*

The measuring and control system builds and tests the steel support axial force servo. A software program and host-implemented measurement and control techniques make this system efficient and accurate (Graph 1,2).

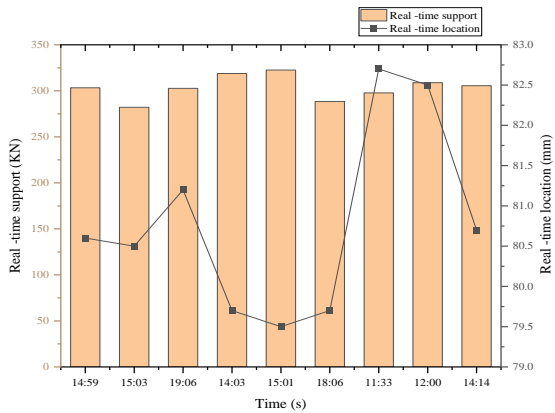

*Graph 1: Servo System Monitoring Data*

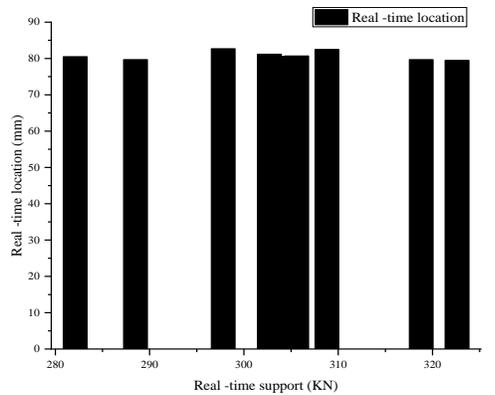

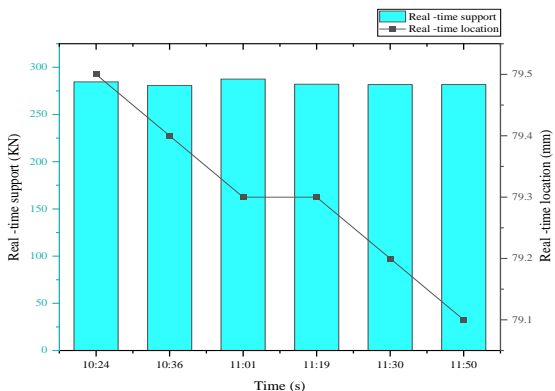

*Graph 2: Servo System Monitoring Data*

CNC displacement/servo sensor pump. A variable frequency motor and valve control hydraulic pressure and displacement closed-loop at this station. Dual displacement/axial force control, automated low-voltage correction, high-voltage warning displacement, real-time monitoring. Protective features function. CNC pumping station Axial force servos are PLC-controlled. Controllers manage systems.

**Discussion**

- A comprehensive design approach will determine the axial force of hydraulic servo steel struts, optimizing deformation regulation in foundation pits.
- Using advanced testing techniques to ensure the control system maintains proper axial force levels and prevents excessive internal forces in the diaphragm wall.
- Investigated unique servo adjustment methods to improve deformation control and stability in deep foundation pit building projects by considering support system interconnectivity.

**Conclusion:**

1. Three-lock systems offer several advantages over double-lock systems, including redundancy, lower capacity operation, and more evenly distributed forces.
2. Three-lock assemblies disperse forces more evenly, reducing stress and enhancing stability. They also distribute load across the head assembly, linking the servo system to the supporting structure, and improving structural integrity.
3. Three-lock systems enhance wear tolerance, simplifying system operation and reducing downtime. They integrate sensors and locking mechanisms, providing lock engagement and wear data for preventive maintenance.

ACKNOWLEDGMENT

The writer would like to express his thanks to his colleagues in working group for help in gathering the data presented here, and in particular, to key laboratory of highway engineering ministry of education in special areas ChangAn University and Sichuan Chuanjiao road and bridge Co.Ltd. for help with sourcing data. The great comments and suggestions from the anonymous reviewers and the Editor are sincerely appreciated.